\documentclass[letterpaper,12pt]{article}   

\usepackage{osajnl2} 
\usepackage{graphics, epsfig}

\begin{document}

\title{Transpose symmetry of the Jones Matrix and topological phases}


\author{\bf Rajendra~Bhandari}
\address{  Raman Research Institute, \\ Bangalore 560 080, India. \\ email: bhandari@rri.res.in}

\begin{abstract}The transmission Jones matrix of an arbitrary stack of reciprocal 
plane parallel plates which has been turned through $180^\circ$ 
about an axis in the plane of the stack is, in an appropriate basis, the transpose of the 
transmission matrix of the unturned slab with a change in the 
sign of the off-diagonal elements. We prove this convention-free 
result for the case where reflection at the interfaces 
can be ignored and use it to devise an experimental scheme to 
separate isotropic and topological phase changes in a reciprocal 
optical medium.

\end{abstract}

\ocis{260.5430, 230.4170, 230.5440, 230.1360}

\maketitle 
\section{Introduction}

Our interest in this problem originated in an attempt to answer the following 
question: Can one find anything in nature or put together anything that could 
act as a ``one-way window" for unpolarized light without the use of magnetic fields, 
i.e. anything that would transmit a different fraction of unpolarized light in one 
direction as compared to the reverse direction. It is generally believed that 
such a possibility is ruled out by fundamental laws of nature.  This implies 
that there must exist a simple relation between the 2 x 2 transmission Jones matrices 
of an arbitrary stack of reciprocal plane parallel plates (films) for forward and backward 
propagation. Several different formulations have been used to deal with this question. 
A convenient and useful way of asking the above question is: If $M$ is the 
transmission Jones matrix of a stack of reciprocal, anisotropic and absorbing 
plane parallel plates with surfaces parallel to the $(\hat x,\hat y)$ plane, expressed in 
the basis of $\hat x$ and $\hat y$ linearly polarized states, with 
the light beam propagating in the $\hat z$ direction and the plate is turned 
through $180^\circ$ about $\hat x$ or $\hat y$, how is the new Jones matrix 
$M^B$ related to $M$ ? The first answer  that comes to mind, following a quick reading of the 
pioneering work of Jones \cite{jones2,jones1} is,  
$M^B=M^T$, where $M^T$ is the transpose of $M$. 
We give below a simple argument to show why this is not true.

Consider the case where the sample is a pure reciprocal optical rotator (for example sugar solution) 
represented by a Jones matrix $R$. Under a $180^{\circ}$ rotation about an axis perpendicular to 
the beam, the rotator is invariant, hence must be represented by the same matrix $R$. But in the basis 
of linearly polarized states, $R$ is given by,

\begin{eqnarray}
 & & R(\phi)=\left( 
  \begin{array}{lr}
{\rm cos}\phi & {\rm sin}\phi \\
-{\rm sin}\phi &{\rm cos}\phi \\
\end{array}
\right) \label{eq:r1}
\end{eqnarray}

\noindent where $\phi$ is the angle of rotation produced by the rotator. 
It is obvious from Eqn.(\ref{eq:r1}) 
that the transpose $R^T(\phi)$ of $R(\phi)$ is not equal to $R(\phi)$. The transmission matrix of the 
turned sample cannot therefore be equal to $R^T(\phi)$.

\section{Statement of the theorem and proof}

The correct answer to the above question is,

\begin{eqnarray}
 {\rm For}~ M=\left( 
  \begin{array}{lr}
m_{11} & m_{12} \\
m_{21} &m_{22}\\
\end{array}
\right), ~  M^B=
\left( 
  \begin{array}{lr}
m_{11} & -m_{21} \\
-m_{12} &m_{22}\\
\end{array}
\right).& & \label{eq:r2}
\end{eqnarray}

\noindent This can be proved through the following steps:\\

 (A) Define an operation ``n-transpose" on any NxN complex matrix $G$ as being one 
under which $G$ goes to $\bar G$ such that

\begin{eqnarray}
\bar G_{ij} = (-1)^{i+j}G_{ji}& & \label{eq:r2a}
\end{eqnarray}

\noindent From Eqn.(\ref{eq:r2a}) it can  easily be shown that

\begin{eqnarray}
& & (\overline{G_1 G_2 ... G_n})^=\bar G_n \bar G_{n-1} ...\bar G_1,\label{eq:r5}
\end{eqnarray}

\noindent where $G_1$, $G_2$ , ..., $G_n$ are NxN complex matrices. \\

 (B) Any arbitrary  Jones matrix $M$ can be written as a product

\begin{eqnarray}
& & M = p~a~ S,\label{eq:r6}
\end{eqnarray}

\noindent where $p={\rm exp}(i\alpha), a= {\rm exp}(-\beta)$;
$\alpha$, $\beta$ being real numbers and $S$ is a 2x2 complex 
matrix with determinant +1, i.e. an element of the group SL(2,C), also called SL(2).\\

(C) It can be shown that $S$ can always be written as a product

\begin{eqnarray} 
S=K_1 K_2 ... K_6, \label{eq:r7}
\end{eqnarray}

\noindent where $K_1, K_2 ... K_6$ are matrices of the form $R$, $L_0$ or $D_0$ where 
$R$ is given by Eqn.(\ref{eq:r1}) and $L_0$, $D_0$ are
given by,

\begin{eqnarray}
 & & L_0(\delta)={\rm diag}~[{\rm exp}(-i\frac{\delta}{2}),~{\rm exp}(i\frac{\delta}{2})], \label{eq:r8}\\
 & & D_0(\gamma)={\rm diag}~[{\rm exp}(-\frac{\gamma}{2}),~{\rm exp}(\frac{\gamma}{2})]. \label{eq:r9}
\end{eqnarray}

\noindent The matrix $L_0(\delta)$ represents a linear retarder with retardation $\delta$ 
and $D_0(\gamma)$ represents an element of dichroism with relative attenuation coefficient $\gamma$, 
the eigenstates of both matrices being linear polarizations along $\hat x$ and $\hat y$. 
Note that the matrices $R(\phi)$, $L_0(\delta)$ and $D_0(\gamma)$ have the property 
\begin{eqnarray}
 & &\bar R(\phi) = R(\phi),~~
\bar L_0(\delta) = L_0(\delta) ~{\rm and}~~\bar D_0(\gamma) = D_0(\gamma). \label{eq:r10}
\end{eqnarray}

We shall prove  statement (C) in the context where 
the 6-parameter group SL(2,C) represents polarization transformations.
Choose the $\hat x$ and $\hat y$ linearly polarized states as the basis states 
for all the unitary 
and nonunitary polarization transformation matrices. 
We first note from the theory of the group SL(2,C), which is homomorphic to the Lorentz Group 
SO(3,1) \cite{grouptheory}, that any element $S$ 
of the group can be written as a product

\begin{eqnarray}
 & & S = NU \label{eq:r11}
\end{eqnarray}

\noindent where $U$ is a unitary matrix, i.e. an element of pure birefringence with 
a pair of orthogonal eigenstates $\mid u>$, $\mid \tilde u>$ and $N$ is a nonunitary matrix, i.e. an 
element of pure dichroism with a pair of orthogonal eigenstates 
$\mid v>$, $\mid \tilde v>$. 
Further, if $U_0$ and $N_0$ represent elements of birefringence and dichroism 
respectively which are diagonal in the chosen basis and with the same eigenvalues 
as $U$ and $N$ respectively, 
there always exist unitary transformations $F$ and $G$ with linearly polarized 
eigenstates  such that $F\mid x>=\mid u>$, $F\mid y>=\mid \tilde u>$, $G\mid x>=\mid v>$
and $G\mid y>=\mid \tilde v>$ so that

\begin{eqnarray}
 & & U = F U_0  F^\dagger ~ {\rm and} ~ N = G N_0 G^\dagger \label{eq:r12}
\end{eqnarray}

\noindent  Note $U_0$ and $N_0$ are matrices of the form $L_0$ and $D_0$ respectively.
 Further, since $F$ and $G$ have linearly polarized eigenstates, 

\begin{eqnarray}
 & & F = L_\psi(\delta_1)=R(\psi)L_0 (\delta_1)R(-\psi)~{\rm and}~ \nonumber\\
 & & G=L_\xi(\delta_2)=R(\xi)L_0 (\delta_2)R(-\xi) \label{eq:r14}
\end{eqnarray}

\noindent where $L_\psi(\delta_1)$ and $L_\xi(\delta_2)$ are linear retarders with retardations 
$\delta_1$ and $\delta_2$ and fast axes making angles $\psi$ and $\xi$ respectively with $\hat y$.  
Eqs. (\ref{eq:r11}), (\ref{eq:r12}) and (\ref{eq:r14}) together constitute a proof of the statement (C). 

(D) Consider now an infinitely thin sample whose Jones matrix in the chosen 
basis is given by a matrix $M$ which can be expressed as in Eqn.(\ref{eq:r6}). 
The matrix $S$ can then be expressed as in Eqn.(\ref{eq:r7})  
where $K_1, K_2 ... K_6$ represent infinitesimal transformations 
given by matrices of the type $R$, $L_0$ or $D_0$. Now rotate the sample 
through $180^\circ$ about the $\hat x$ or $\hat y$ axis. The Jones matrix 
$S^B$ of the reversed sequence of elements is given by 

\begin{eqnarray}
 & & S^B = K_6^B. K_5^B...K_1^B \label{eq:r15}
\end{eqnarray}

\noindent where $K_n^B$ is the matrix of the reversed version of  $K_n$. 
Since each of the elements  $R$, $L_0$ and $D_0$ 
is physically invariant under such a rotation, $K_n^B=K_n$ for all $n$.
Hence 
\begin{eqnarray}
 & & S^B = {K_6}. {K_5}...{K_1} \label{eq:r16}
\end{eqnarray}

\noindent Then, from Eqns.(\ref{eq:r10}) and (\ref{eq:r5}), we have,

\begin{eqnarray}
 & & S^B = {\bar K_6}. {\bar K_5}...{\bar K_1}=\bar S \label{eq:r17}
\end{eqnarray}

\noindent Since the isotropic factors of  $M$ commute with all 
operations, we therefore have,
 
\begin{eqnarray}
 & & M^B =\bar M \label{eq:r18}
\end{eqnarray}

(E) We now make the only assumption in this proof which is necessary in view of 
possible absorption in the sample. We assume that if two 
infinitesimally thin samples have the same transmission Jones matrix $M$  for 
forward propagation they must have, in the absence of non-reciprocal effects, 
the same transmission Jones matrix $M^B$ for
reverse propagation.  Eqn.(\ref{eq:r18}) therefore holds for the original infinitesimal sample. 
Now since a sample of finite thickness can be looked upon as an infinite 
sequence of infinitesimally thin samples and since the above sequence of arguments can 
be repeated for such a sequence, it follows that Eqn.(\ref{eq:r18}) holds for a finite sample,
which is our main result. 

The reason for the difference between our result and that of Jones is that 
in  \cite{jones1} the sample 
is kept fixed and the  beam is reversed. This 
requires, in addition to an effective $180^\circ$ rotation of the sample about $\hat y$, 
a  convention for the relative phase between the basis states of the  matrices 
for the two directions of propagation. In the convention used in  \cite{jones1}, 
if the rotation is about the $\hat y$ axis, there is a $\pi$ phase differencce 
between the two $\hat x$ polarized basis states and none between the two 
$\hat y$ polarized ones. This implies a relative phase shift of $\pi$ 
between the $\hat x$ and $\hat y$ polarizations in switching the direction of 
propagation,  corresponding to a unitary transformation by means of  $\sigma_3$=diag[1,-1]. 
Indeed it can be verified that  
$\bar M$ and $M^T$ are related by

\begin{eqnarray}
 & & M^T = \sigma_3~ {\bar M}~\bar {\sigma_3}^\dag. \label{eq:r24}
\end{eqnarray}

\noindent Jones' result  is therefore consistent with ours. In our formulation however, 
the direction of propagation remains fixed. We therefore use the same set of basis 
states with the same phases for the forward and the backward matrices which is the 
simple and natural thing to do and one does not require a phase convention. 
Eqn. (\ref{eq:r18}) can be verified in a 
sequence of simple null interference experiments 
which will be described  elsewhere. From Eqns.(\ref{eq:r18}) and  (\ref{eq:r2a}) 
it follows that for 
an incident unit intensity unpolarized beam, the intensity transmitted by the reversed 
sample $I^B_T =\frac{1}{2} Tr( {M^B}^\dag M^B)$ is the same as that 
transmitted by the original sample i.e. $I_T = \frac{1}{2}Tr( M^\dag M)$.

\section{\bf Separation of isotropic and topological phases}

It is now well known \cite{martinelli,vandeventer} that if a beam of polarized 
light passes through an arbitrary 
reciprocal medium, then through a $45^\circ$ Faraday rotator and is then 
reflected normally off an isotropic plane mirror so that it retraces its 
path through the medium, 
the effect of any reciprocal birefringence or dichroism 
in the medium is cancelled.  Eqn.(\ref{eq:r18}) together with the method 
of analysis described in  
\cite{decomposition} provides a very compact proof of this result.

In Fig.(1a) let $M$ stand for any reciprocal optical medium, $FR(45)$  
for a Faraday rotator that rotates the polarization about the beam axis through 
$45^\circ$ in real space and the ``mirror" stand for any isotropic reflector 
placed normal to the beam. Let us  choose the $\hat z$  axis as the 
propagation direction and the $\hat x,\hat z$ plane to be 
the reflection plane \cite{footnote}. 
The medium $M$ can be decomposed 
as in Eqn.(\ref{eq:r6}) into (i) an isotropic refractive part $p$, (ii) an isotropic absorption part $a$ and 
(iii) $S$, an element of the group SL(2,C). Following \cite{decomposition}, the  reflection at the mirror 
is replaced by a halfwave plate $H_0$
with its fast axis along  $\hat y$,   
$FR(45)$   and $M$ 
encountered on the reverse passage are replaced by equivalent 
elements which have been rotated about $\hat y$ through $180^\circ$ and 
placed in the forward path of the beam. This yields an equivalent optical circuit, shown in Fig.(1b),
with a round trip Jones matrix $M^{(\rm rt)}$  given by  

\begin{eqnarray}
 M^{(\rm rt)}= a ~p~\bar S~ R(-45)~H_0R(45)~ S~ p~ a =a^2 p^2\bar S~ H_{45}~ S  & & \label{eq:r38}
\end{eqnarray}

\noindent where we have used Eqn.(\ref{eq:r17})  to represent the 
reversed medium by $\bar S$.  The matrix $H_{45}$  has the following form:

\begin{eqnarray}
 & & H_{45}=-i\left( 
  \begin{array}{lr}
0 & 1 \\
1 &0 \\
\end{array}
\right) \label{eq:r39}
\end{eqnarray}

\noindent The crucial part of the proof consists of the following identity which 
can be easily verified :

\begin{eqnarray}
\bar SH_{45} S=H_{45}.& & \label{eq:r40}
\end{eqnarray}

\noindent where $S$ is any 2x2 complex matrix
with determinant +1 and $\bar S$ is defined by Eqn.(\ref{eq:r2a}). Eqns.(\ref{eq:r38}) and 
(\ref{eq:r40}) then give

\begin{eqnarray}
  M^{(\rm rt)} = a^2 p^2H_{45}& & \label{eq:r41}
\end{eqnarray}

\noindent As shown in Figs.(1b) and (1c), the polarization evolution described by  $M^{(\rm rt)}$  is 
followed by a rotation of the beam about $\hat y$ through $180^\circ$ \cite{decomposition}.
Eqn.(\ref{eq:r41}) shows that the polarization state of the return beam  is independent $S$, i.e. 
of any parameter representing 
birefringence and dichroism in the medium.
Since such a cancellation is well supported by experiment and since an equation like (\ref{eq:r40}) 
is not obtained if $S^T$ were used in place of $\bar S$ we consider 
Eqn.(\ref{eq:r40}) an indirect experimental support for our Eqns.(\ref{eq:r17}) and (\ref{eq:r18}).\\

\begin{figure*}
\centerline{
\epsfxsize=0.7\textwidth
\epsfbox{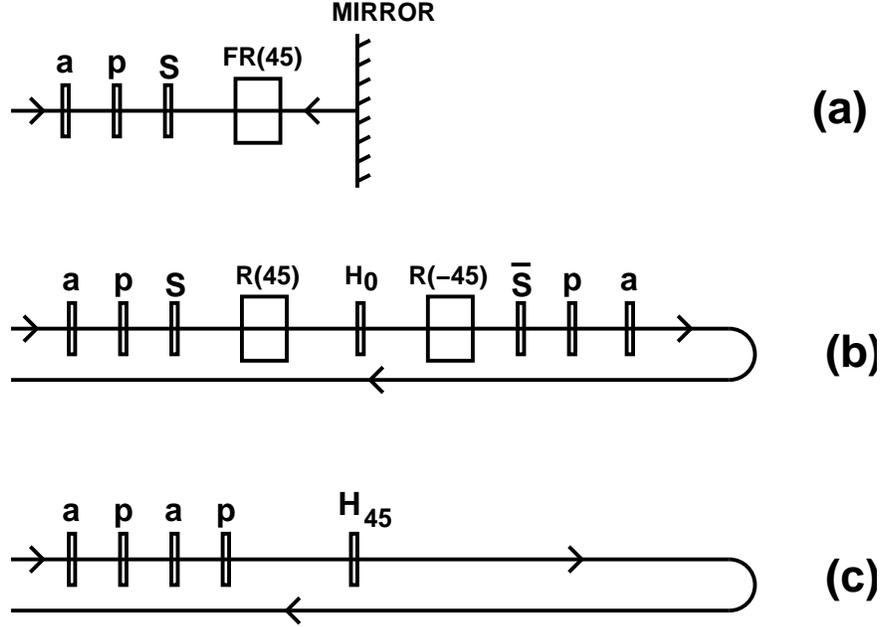}
}
\caption{(a) Double passage through a medium by means of the  $45^\circ$ Faraday 
rotator reflector.
(b) Equivalent optical circuit.  (c) The reduced equivalent circuit.}
\label{fig.1}
\end{figure*}

Eqn. (\ref{eq:r41}) has another interesting consequence. While all phase changes that arise 
due to changes in the SL(2,C) part of the medium
(geometric phases, Pancharatnam phases etc. \cite{rbreview}) are cancelled,  the isotropic phase factor $p$ 
is not cancelled by double passage.
This provides a method 
for separately determining isotropic and topological phase changes in a medium. 
An interference experiment in which the beam passes through the medium only once 
measures the total phase shift $\Delta \phi_1=\phi_{\rm iso}+\phi_{\rm topo}$ whereas 
an interference experiment with double passage by means of a 
Faraday mirror  measures $\Delta \phi_2=2\phi_{\rm iso}$. From  $\Delta \phi_1$ 
and $\Delta \phi_2$ the isotropic phase shift 
$\phi_{\rm iso}$ and the topological phase shift $\phi_{\rm topo}$ can be determined 
separately. In fact as shown in Fig.(2), the two 
interferometers  can be combined in a single setup with a 
common arm containing the experimental medium $M$. The quantities $\Delta \phi_1$ 
and $\Delta \phi_2$ can then be measured simultaneously. 

\begin{figure*}
\centerline{
\epsfxsize=0.7\textwidth
\epsfbox{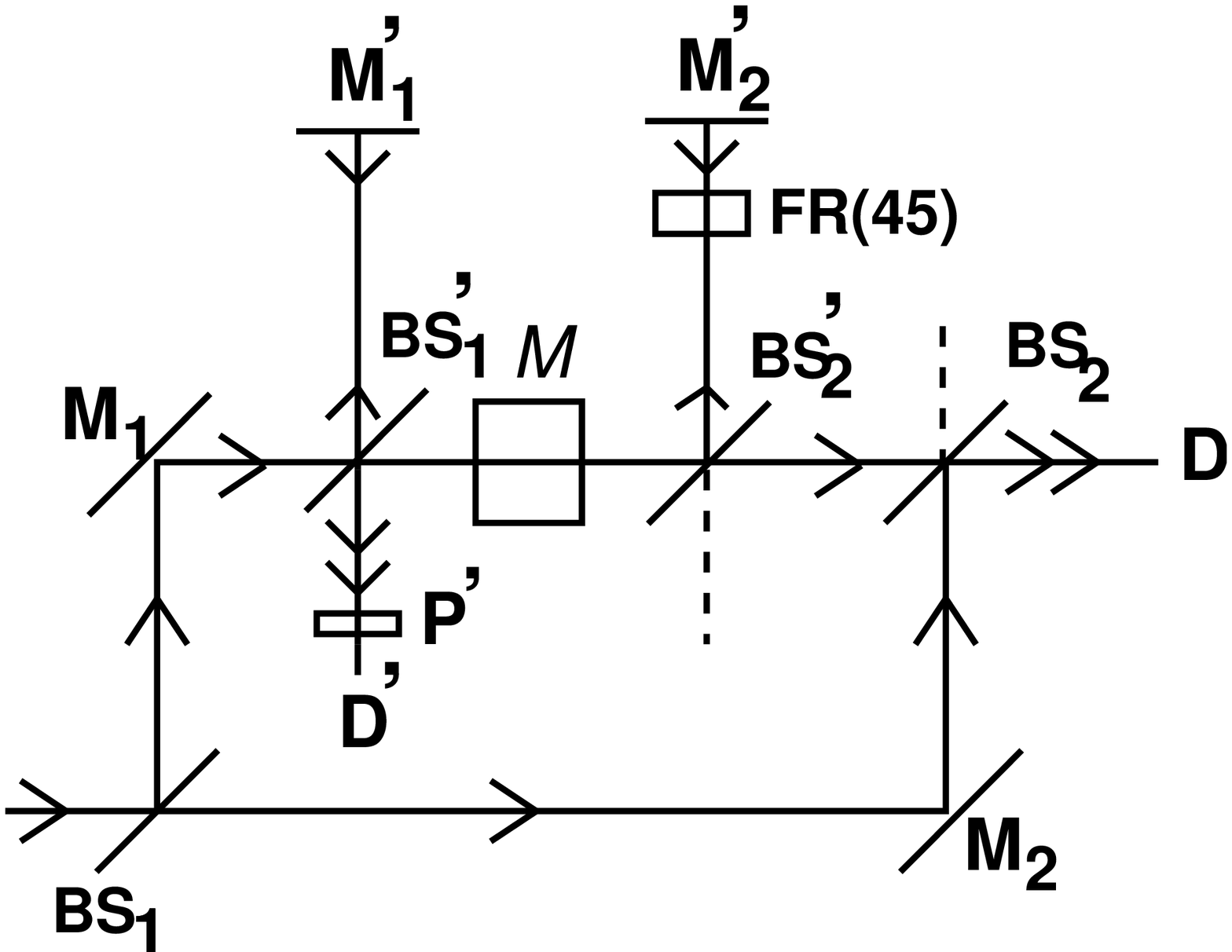}
}
\caption{An interference setup to determine simultaneously the isotropic and topological 
phase shifts in a reciprocal optical medium $M$. The beam splitters $\rm BS_1'$, $\rm BS_2'$, 
the mirrors $\rm M_1'$, $\rm M_2'$, the polarizer P' and the fringe detector $\rm D'$ comprise a double pass Michelson interferometer 
with a $45^\circ$ Faraday rotator and 
$\rm BS_1$, $\rm BS_2$, $\rm M_1$, $\rm M_2$, $\rm D$ comprise a single-pass Mach-Zhender interferometer.}
\label{fig.2}
\end{figure*}

\section{Acknowledgements}

It is a pleasure to thank Bala Iyer for several helpful discussions.

\newpage


\begin{thebibliography}{4pism}
\bibitem{footnote}
It can be  shown  that the final state of the beam as 
obtained by this method is independent of the choice of the reflection plane.
\end{thebibliography}
\end{document}